\documentclass[letterpaper]{article} 
\usepackage{aaai24}  
\usepackage{times}  
\usepackage{helvet}  
\usepackage{courier}  
\usepackage[hyphens]{url}  
\usepackage{graphicx} 
\usepackage{natbib}  
\usepackage{caption} 
\usepackage{algorithm}
\usepackage{algorithmic}
\usepackage{amsmath}
\usepackage{tabularx}

\usepackage{newfloat}
\usepackage{listings}
\DeclareCaptionStyle{ruled}{labelfont=normalfont,labelsep=colon,strut=off} 
\floatstyle{ruled}
\newfloat{listing}{tb}{lst}{}
\floatname{listing}{Listing}
\title{Migrating a Job Search Relevance Function}
\author{
    \equalcontrib{Bennett Mountain},
    \equalcontrib{Gabriel Womark},
    Ritvik Kharkar
}
\affiliations{
    Ziprecruiter, Inc\\
    604 Arizona Ave\\
    Santa Monica, California 90401 USA\\
    \{bennettm, gabrielw, ritvikk\}@ziprecruiter.com

}

\begin{document}
\frenchspacing

\maketitle

\begin{abstract}
In this paper, we describe the migration of a homebrewed C++ search engine to OpenSearch, aimed at preserving and improving search performance with minimal impact on business metrics. To facilitate the migration, we froze our job corpus and executed queries in low inventory locations to capture a representative mixture of high- and low-quality search results. These query-job pairs were labeled by crowd-sourced annotators using a custom rubric designed to reflect relevance and user satisfaction. Leveraging Bayesian optimization, we fine-tuned a new retrieval algorithm on OpenSearch, replicating key components of the original engine's logic while introducing new functionality where necessary. Through extensive online testing, we demonstrated that the new system performed on par with the original, showing improvements in specific engagement metrics, with negligible effects on revenue.
\end{abstract}

\section{Introduction}
\label{sec:intro}
Job search is integral to everyone's journey of finding a new job. What was once a simple act of peering up and down the classified section containing only 10s of jobs in a newspaper is now a complex act of discovering your next role by sifting out 10s of millions of jobs in a search engine. In 2014, ZipRecruiter entered into the job search scene with a C++ based search engine that though admirably served its purpose, has since become outdated. Over the past 10 years, search engine technology has rapidly evolved with the rise of artificial intelligence and vector search taking up the lion's share of the advancement with open-source search engines leading the charge. That, coupled with our C++ search engine growing into a monolith, led us to make the decision to migrate to OpenSearch, an open-source fork of ElasticSearch maintained by AWS.  

The C++ search engine's lexical relevance function was not directly portable to OpenSearch due to the structure of its query DSL, requiring significant work to create a new relevance function that adhered to OpenSearch's constraints. Though we were able to take inspiration from the old relevance function, this inability to directly port it over required us to create a new functional form for our relevance function and smartly tune field-boosts. Our new relevance function not only matched the performance of the C++ search engine's relevance function, it exceeded the C++ search engine's relevance function on all key metrics.

\section{Methodology}
There are two key facets of our relevance function. The first is the \textit{functional form}, or the overall structure that a) determines what fields from each job are used for retrieval and ranking and b) how BM25 scores \cite{article} from different fields are utilized and combined to determine the overall score of a job. The second is the \textit{field boosts} that are used to weight how much we care about query token matches in each field of a job by scaling up or down the BM25 score of a field match for a query token. Though these two factors can be tweaked independently, we achieved the most success when modifying them in tandem.

Crafting the functional form of the relevance function and finding the optimal field boost values took up a significant amount of time. At a high level, here's how we accomplished that: 

\begin{enumerate}
    \item Create definitions of search relevance grades to represent the degree of relevance for a query and a job.  
    \item Create a labeled data set $D$ of (query-job-relevance label) triplets
    \item Create a functional form of a relevance function inspired by our C++ relevance function
    \item Use $D$ to evaluate the functional form and tune OpenSearch Query DSL boosts and other hyperparameters in the relevance function
\end{enumerate}

\subsection{The C++ Search Engine's Relevance Function}
The meticulously crafted and tuned relevance function from our C++ search engine was TF-IDF based \cite{roul2014webdocumentclusteringranking} with a semantic scoring element. Given a job search query, the relevance function outputs a score from 0-1 that seeks to capture how \textit{relevant} (see Section 1.2) a job is to a query. At its core, a job on ZipRecruiter contains a title, description, and company name. For each job, the C++ search engine processed these raw fields using rewrite and stemming/lemmatization rules (analyzer rules), generating new fields. It applied the same analyzer rules to the query, using both as inputs for the relevance function: 
\[
rel\_score = f(analyzed\_query, analyzed\_doc\_fields)
\]
This $rel\_score$ score, combined with other factors such as how close a job is to a job seeker's search location (geo) along with how newly posted a job is (freshness) ultimately determines whether or not it will be retrieved by our search engine and its ranking position. However, we will not focus on geo and freshness in this paper, as we were able to directly port over those factors to the new relevance function.

\subsection{Defining Relevance}
 At ZipRecruiter, our mission is to actively connect job seekers to their next great opportunity. Thus, we define query-job relevance as ``how likely would a job seeker who searches for this query be to apply to this job?". It's worth mentioning that to completely capture how likely someone is to apply to a job, we would need to know their preferences on salary, location, how new a job is, etc. These factors we refer to as \textit{user} relevance. On the other hand, \textit{topical} relevance contains factors such as how well the query's intent matches the responsibilities, requirements, and skills of a job. Since we could directly port over our scaling factors for user relevance from our C++ search engine's relevance function, in this migration we were only concerned with assessing \textbf{topical relevance}. Moving forward, when we refer to ``relevance", we're really referring to topical relevance. To measure the relevance of a query-job pair, we defined five grades of relevance as shown in Table \ref{table:job_relevance}. Human labelers were prompted with specific instructions to ignore user relevance and were tasked with labeling tens of thousands query-job pairs according to the below relevance definitions. These relevance labels were the keystone for creating our new relevance function, as discussed in detail in Section \ref{sec:offline_eval}.

\begin{table}[h!]
\centering
\renewcommand{\arraystretch}{1.3} 
\begin{tabular}{|c|p{2.5in}|} 
 \hline
\textbf{Grade} & \textbf{Description} \\ 
 \hline\hline
 0 & This is far from what I'm looking for. There's no way I would apply—\textbf{extremely irrelevant}. \\ 
 \hline
 1 & This job doesn't align well with my search query. There's a very small chance I would apply—\textbf{irrelevant}. \\
 \hline
 2 & I can see why I got this result. I might apply to it —\textbf{somewhat relevant}. \\
 \hline
 3 & This job isn't the best match, but still a good one. I would apply to it —\textbf{relevant}. \\
 \hline
 4 & This job is a perfect match. I would put in a lot of effort to apply to this job—\textbf{extremely relevant}. \\ 
 \hline
\end{tabular}
\caption{Job Relevance Grades}
\label{table:job_relevance}
\end{table}

When working with metrics that required labels to be binarized as either relevant or irrelevant, we set our relevance threshold at 3 such that any job with a relevance label greater than or equal to that threshold was deemed relevant and irrelevant otherwise.

\subsection{Collecting Relevance Labels}
To obtain our dataset $D$ of query-job relevance labels, we first generated a candidate set of query-location pairs using a set of heuristics described below. Next, we ran searches using the C++ search engine against a static snapshot of our search index to generate result sets containing query-job pairs. We then sent those query-job pairs to human labelers to obtain graded relevance labels. Our goal was to create a balanced dataset of relevant and irrelevant jobs so that 1) our new relevance function could learn to identify what makes a job relevant or irrelevant for a given query, and 2) future ML models could be trained on it.

\subsubsection{Queries}
 The job search query space is vast and complex, containing queries as simple as ``full time" to queries as complex as ``ziprecruiter senior remote python machine learning engineer". We wanted to bias towards head queries that users most frequently search for but still wanted to include queries from our torso in order to be representative of our traffic. To do this, we first took our top 2,000 queries (representing 80\% percent of our traffic), manually combed through them to remove queries whose relevance labeling tasks would be considered trivial. For example, jobs for the query ``full time" could trivially be labeled based on whether the job offered full time employment. We then obtained named entity recognition (NER) tags for each query\footnote{As an example, ``ziprecruiter senior remote python machine learning engineer" is tagged as [B-company, B-seniority, B-job\_type, B-area\_of\_interest\_specialty, B-job\_title, I-job\_title, I-job\_title]} using our in-house job search query tagger inspired by \cite{cheng2020endtoendsolutionnamedentity}, and further categorized queries by their number of tokens. With our queries categorized, we drew stratified random samples from each category to ensure that our query set wasn't overly biased towards a single category (such as unigram job titles). The list of query tags can be found below in Table \ref{table:ner_tags}.

\begin{table}[h!]
\centering
\renewcommand{\arraystretch}{1.3} 
\begin{tabularx}{\linewidth}{|c|X|X|} 
 \hline
 \multicolumn{3}{|c|}{\textbf{NER tags}} \\
 \hline
 \hline
  workplace & job\_type & job\_title \\
 \hline
 generic\_title\_implied\_seniority & seniority & company\\
 \hline
 area\_of\_interest\_specialty & other &\\
 \hline
\end{tabularx}
\caption{Tags from our Query NER Tagger}
\label{table:ner_tags}
\end{table}

\subsubsection{Locations}
The job search location space is similarly vast, with job seekers searching for locations as broad as ``remote" and ``USA" to as specific as ``Manchester, Vermont" and ``Tribeca". Locations in which have a large inventory of jobs present several issues for collecting relevance judgments. Searches in these locations generate a prohibitively large number of results to be labeled with our allocated labeling budget. Furthermore, as these locations have a large inventory of jobs in general, they also often have a larger inventory of relevant jobs to serve in search results, preventing us from achieving our desired balance of irrelevant and relevant results for relevance tuning. To avoid these problems, we needed to pare down our possible location space of over 100,000 unique areas (states, cities, towns, neighborhoods, etc) in the US to a few hundred that provided us with what we needed. To do this, we used an area's population as a proxy via wikidata \cite{wikidata_population}, randomly selecting a few hundred whose populations were between 10,000 and 30,000.

\subsubsection{Query-Location Pairs}
After collecting a candidate set of query-location pairs, we executed searches using the C++ search engine for each pair on our static snapshot of our job index, obtaining tens of thousands of candidate result sets. From these, we only selected those that had our desired distribution of relevant and irrelevant jobs where relevance was estimated using C++ search engine's relevance function score and between 10-100 results. If multiple candidate locations were obtained for a query, we selected the location with the fewest number of duplicate jobs. When all was said and done, we had several hundred search result sets generated from our query-location pairs, containing thousands of query-job pairs. We sent these query-job pairs to a third-party labeling service to get relevance grades, ultimately resulting in the creation of the dataset $D$ as mentioned in Section \ref{sec:intro}.

\section{Offline Evaluation}
\label{sec:offline_eval}
Getting relevance grades for query job pairs is a common practice in search relevance used to get a quantitative assessment of how good a search engine is \cite{moniz2016datadrivenrelevancejudgmentsranking}. With relevance grades, we can calculate metrics like precision, recall, and NDCG to get a holistic picture of our search engine's performance. NDCG is a commonly used metric used to measure how well-ordered a result set is compared to its ideal ordering \cite{wang2013theoreticalanalysisndcgtype}. For us, our ideal ordering of a result set is when it's ordered by relevance grades descending: with the most relevant results at the top and the most irrelevant results at the bottom. This directly captures what searches expect: the results they're looking for at the top with less relevant results appearing as they scroll down. During our offline evaluation, each ``example" in our dataset was a query-location pair (ie an entire result set) as opposed to a query-job pair, as the  above mentioned metrics are calculated on a per-result set basis as opposed to a per query-job pair basis. Our baseline was to match the C++ search engine's relevance function on recall; our primary goal was to beat it on NDCG@5. We chose 5 as our cutoff based on user research of what job seekers care about the most on a SERP (search engine result page).

\subsection{Crafting the Functional Form}
Starting with the C++ search engine's relevance function as inspiration, we crafted an initial functional form with the baseline goal of matching the C++ search engine's recall. Recall crucially measures the proportion of relevant jobs returned out of the possible relevant jobs. Since we used the C++ search engine to create our relevance label data set, this meant that our new relevance function had to return every single relevant job that the C++ engine did. Starting with the same way of retrieving jobs based on matching query tokens in certain fields, we were able to match the C++ search engine's recall by simply retrieving all jobs that had a lexical match. Further, we took advantage of a new field our C++ search engine didn't have access to during retrieval in order to retrieve even more jobs. 

In lexical scoring, the strength of a lexical match can be quantified in different ways such as how important each individual token match is, how many tokens of the query were matched, and what fields in the job the tokens were matched in. The C++ search engine's relevance function used a combined term and field-centric approach to lexical scoring, which was not possible in OpenSearch at the time of our development. Forced to deviate from the functional form of our previous relevance function, we tested out both term-centric and field-centric approaches in our new relevance function as well as combinations of the two in different variations. 

\subsection{Tuning}
The OpenSearch query DSL provides search relevance engineers with the ability to weight how much they care about lexical matches by scaling the BM25 score for each job field as well as different other matching clauses. In our new relevance function, we initially had over 10 field boosts that we needed to tune. Though we began with attempting to tune each boost by hand to build intuition, we quickly ran into a problem: when some segments of queries would benefit from a change in a boost, others would suffer. With a dataset containing dozens of query segments and a thousand query-location pairs, having over that many boosts to tune by hand is intractable. To solve this, we employed Bayesian optimization \cite{frazier2018tutorialbayesianoptimization}. Bayesian optimization is effective for finding the optimal values in a multi-dimensional search space with a computationally expensive objective function. We optimized for NDCG@5, and each iteration evaluated all 1,000 query-location pairs by executing search requests for each before calculating the average NDCG@5 across the result sets.

\begin{figure}[H]
    \centering
    \includegraphics[scale=0.12]{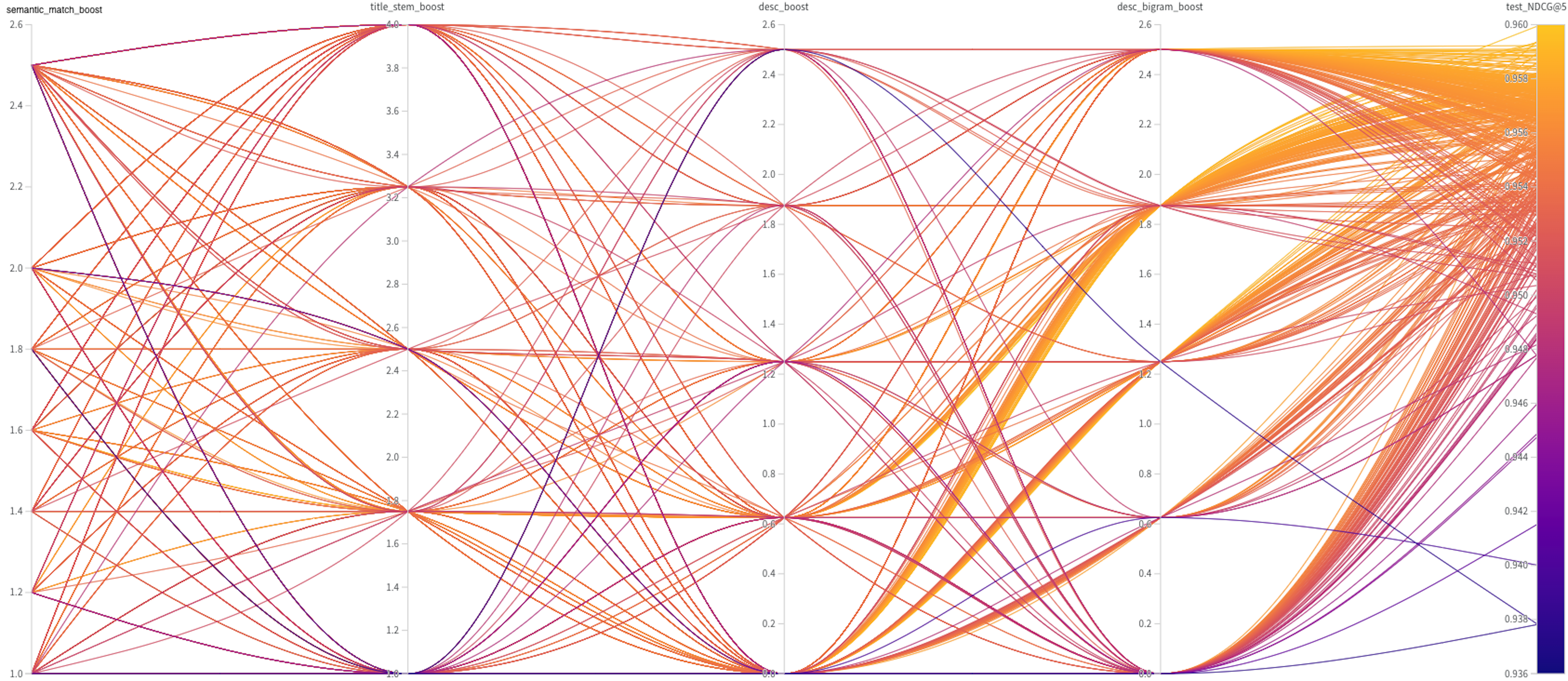}
    \caption{An example sweep of Bayesian optimization}
    \label{relevance_tuning_bayes_opt_example}
\end{figure} 

Figure \ref{relevance_tuning_bayes_opt_example} shows one of our Bayesian optimization runs with a few example boosts. Here, the more yellow a line is in the parallel coordinates plot, the higher the test set NDCG@5. In this, we can especially see how over time the optimizer learns  which parameter values to exploit as shown by more concentrated yellow lines for particular values.

\subsection{The Overall Process}
We started initially by trying to optimize over our training dataset, tuning over 10 boosts each with continuous values. However, we immediately found our optimizer was struggling to learn. In Bayesian optimization, the complexity of a parameter space is dictated by its dimensionality and the possible values that each parameter could take on. To remedy this, we employed two key methods. First, we decided to heavily pare down the dimensionality by setting fixed boost values for fields we had strong intuition and conviction on: this intuition was built through our aforementioned process of manual tuning. Secondly, we gave the optimizer access to less then 10 possible discrete values for each parameter rather than a continuous space.

At the same time, we decided to start with a single segment of our query space to further reduce the difficulty for the optimizer. The intuition here is that queries in the same segment (such as unigram job title queries of ``nurse") \textit{should} should have a similar optimal set of boosts as one another and exhibit similar behavior when changing boosts. Intuitively, if a job title token in a query appears in the title field of a job, the job is likely relevant so you want it to be higher up in your ranking.

\begin{figure}[H]
    \centering
    \includegraphics[scale=0.19]{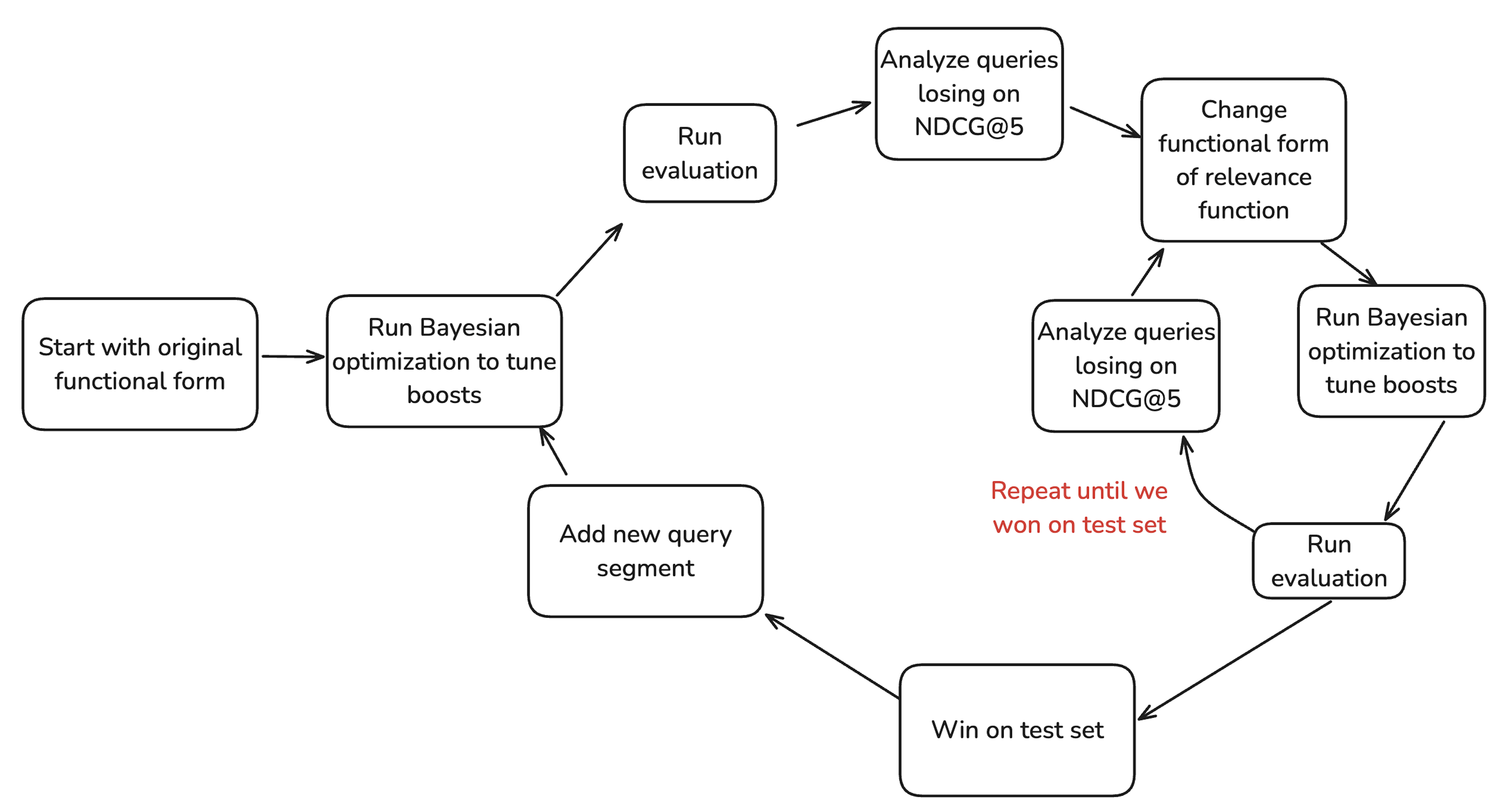}
    \caption{The overall process of relevance tuning}
    \label{relevance_tuning_process}
\end{figure}

However, sometimes we still weren't able to beat the C++ search engine's relevance function. Here, we dug into the results from our optimization, looking at queries that we were losing the most on, crucially analyzing the factors that went into the final relevance score for each result in a result set. This qualitative evaluation not only allowed us to build more intuition around what the optimal set of boosts should look like for each query and segment, but most importantly was the catalyst for coming up with novel ideas to change the functional form of our relevance function. With each change in our functional form came a new run of our optimizer. We then repeated this process until we beat the C++ relevance function on our test set NDCG@5\footnote{It's also worth noting here that manually correcting bad relevance labels from our crowd-sourced annotators also provided us with increases in our evaluation metrics, as individual labels are often inconsistent among annotators even for the same data}. Then and only then did we add a new query segment and repeat again until all segments were covered. An overview of this process is outlined in Figure \ref{relevance_tuning_process}.

This process also led us to come up with new rewrite and stemming rules for our tokenizer, ideas for manual treatments of particularly difficult queries, and ways to improve our semantic query classifier. The biggest unlocks for the functional form of the new relevance function were penalizing jobs where not all query tokens were present in its fields and boosting jobs where all query tokens appeared in the title.

\subsubsection{Tuning BM25 parameters}
OpenSearch also provides search relevance engineers with the ability to tune the BM25 parameters $b$ and $k$ for each field in the index. $b$ is the field length penalty, meaning jobs with longer fields are penalized. $k$ is the term saturation, which controls how many occurrences a term must appear in a field in order to be considered relevant.

With our set of field boosts held constant, we tuned these for a few fields resulting in a lift of 0.5\% in NDCG@5: a non-negligible increase for minimal effort.

\begin{figure}[H]
    \centering
    \includegraphics[scale=0.35]{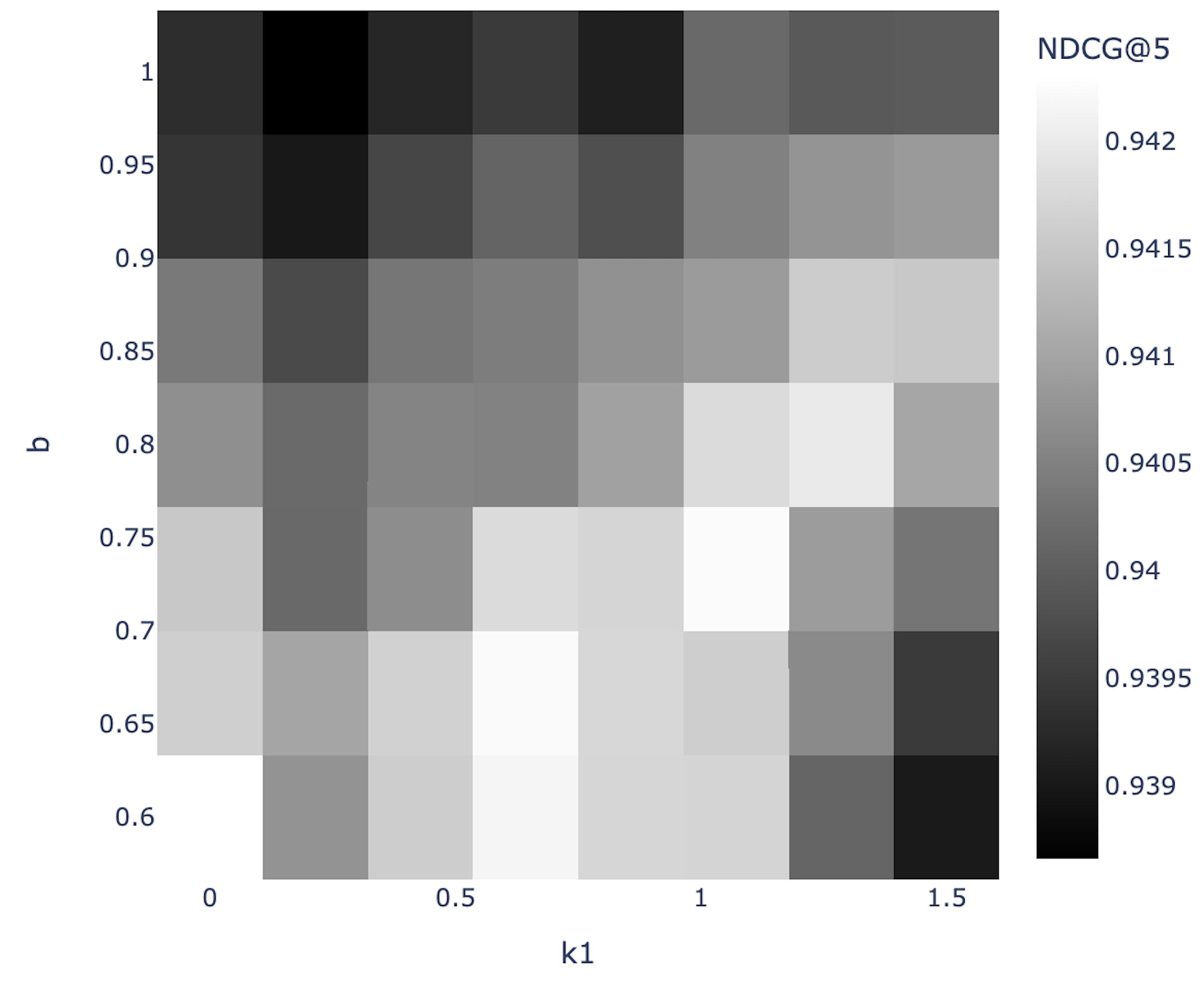}
    \caption{Tuning $b$ and $k$ on an example field in our index with all other boosts held constant}
    \label{ndcg_heatmap}
\end{figure} 

As can be seen in Figure \ref{ndcg_heatmap} showing an example field we tuned $b$ and $k$ on, they were not independent in their effect on NDCG@5. We found this lack of independence to be the case in every field, but found that different fields exhibited different correlations of NDCG@5 with $b$ and $k$. This is backed up by blog posts from OpenSearch and Elasticsearch.

\section{Online Evaluation}
After obtaining a sufficient functional form and set of boosts, we conducted an A/B test to evaluate our solution's performance online. The control variant used the existing C++ search engine and the test variant used the OpenSearch based engine with the aforementioned relevance function and boosts. Normally, after the top 1000 jobs for a query are retrieved by the engine, a separate reranking layer is run before result set pagination to balance relevance of results with monetization based on user engagement. Since we wanted a clear measurement of our new search engine's impact on relevance we disabled the reranking layer for the duration of the experiment. 

In order to determine the success of the new search engine, we relied on measurements of user engagement. More specifically, we relied on an engagement signal called an apply start. Our search engine serves results to users in a two-pane user interface. The left pane exposes a list of jobs for job seekers to engage with, while the right pane shows expanded details for a job that has been clicked on in the left pane. Each query entered generates a set of jobs called an \textit{impression set}. Additionally, the right pane includes a button which allows job seekers to submit an application for the job listing. Clicks on this button are called apply starts. Our hypothesis was that if the new search engine was serving more relevant results, we would see an increase in apply starts per query entered per user.  The metric used to measure this hypothesis is called right pane impression set CTR.

One problematic aspect of this metric is that it's highly dependent on the number of queries users enter. However, there were no significant differences between test and control in the number of queries entered per user, making it suitable for us to evaluate this hypothesis. Additionally, we measured the number of apply starts per user, regardless of the number of queries entered per user, to account for this problem. 

\begin{table}[h!]
\centering
\renewcommand{\arraystretch}{1.3} 
\begin{tabularx}{\linewidth}{|c|>{\centering\arraybackslash}X|} 
 \hline
\textbf{Metric} & \textbf{\%-Relative Lift (99\% CI)} \\ 
 \hline\hline
 right pane impression set CTR & [3.67\%, 10.68\%] \\ 
 \hline
 apply starts per user & [0.45\%, 13.66\%] \\
 \hline
\end{tabularx}
\caption{Online Performance}
\label{table:online_performance}
\end{table}

Table \ref{table:online_performance} shows the results for our new engine when compared to control on the key metrics we previously identified. As we can see, the new engine showed statistically significant improvements upon both metrics, validating our hypothesis which was based on the offline evaluation of our new relevance function.

\section{Future Work}
As we introduce new methods of search engine retrieval or make modifications to our relevance function we will likely need to employ many of the techniques described in this paper to measure relevance improvements and tune new parameters and boosts. For example, our current engine mainly relies on lexical matching technology to retrieve relevant jobs for a user's query, but we'd like to leverage transformer models and dense vector search technology to retrieve relevant jobs that could not possibly be retrieved by the current lexical search engine. In this future we'd likely combine retrieved jobs from both lexical and dense vector search engines and we will likely need to tune parameters that control how much a particular engine's score contributes to the overall score for a job. We could collect query-job pairs, get them labeled, and repeat the Bayesian optimization process to identify the best values for these parameters. 

Additionally, we recognized that while we were able to measure our new relevance function's ability to match the recall of the existing search engine, we never conclusively proved that the new relevance function was capable of retrieving \textit{relevant} results that weren't able to be retrieved by the existing engine. This was due to the fact that we only collected relevance judgments for query-job pairs retrieved by the existing engine. To properly measure improvements on recall in the future, we won't merely collect judgments using our existing relevance function and search engine, but also collect judgments from any new system we desire to migrate to. This will allow us to effectively measure whether the new system was capable of retrieving jobs that the old system could not.

\section{Conclusion}
In this paper, we discussed our experience  migrating from an existing search engine to a new one, specifically focusing on preserving the relevance of search results. While several resources lay out the high-level steps to perform such a migration, our paper delves into an actual implementation of these steps, providing useful, detailed insights that are often not discussed in these resources. We provided a definition of relevance in job search and devised unique strategies for collecting data to evaluate a search engine. We proposed a novel approach for using a well-known machine learning based algorithm to tune parameters for a search relevance function. Lastly, we showed that this approach to tuning and evaluating relevance functions can not only maintain the relevance of search results in an online setting, but actually improve upon it. We share this knowledge with the hopes that others may apply it when attempting to migrate their search engines or simply make improvements upon their existing relevance functions.   

\bibliography{migrating_a_search_relevance_function}

\end{document}